\documentclass[twocolumn,showpacs,prl,superscriptaddress]{revtex4}

\usepackage{graphicx}
\usepackage{amsmath}
\usepackage{amssymb}

\begin{document}

\title{Europium nitride: A novel diluted magnetic semiconductor}

\author{Do~Le~Binh}
  \affiliation{The MacDiarmid Institute for Advanced Materials and Nanotechnology, School of Chemical and Physical Sciences, Victoria University of Wellington, PO Box 600, Wellington 6140, New Zealand}
\author{B.~J.~Ruck} \email{ben.ruck@vuw.ac.nz}
  \affiliation{The MacDiarmid Institute for Advanced Materials and Nanotechnology, School of Chemical and Physical Sciences, Victoria University of Wellington, PO Box 600, Wellington 6140, New Zealand}
\author{F.~Natali}
  \affiliation{The MacDiarmid Institute for Advanced Materials and Nanotechnology, School of Chemical and Physical Sciences, Victoria University of Wellington, PO Box 600, Wellington 6140, New Zealand}
\author{H.~Warring}
  \affiliation{The MacDiarmid Institute for Advanced Materials and Nanotechnology, School of Chemical and Physical Sciences, Victoria University of Wellington, PO Box 600, Wellington 6140, New Zealand}
\author{H.~J.~Trodahl}
  \affiliation{The MacDiarmid Institute for Advanced Materials and Nanotechnology, School of Chemical and Physical Sciences, Victoria University of Wellington, PO Box 600, Wellington 6140, New Zealand}
\author{E.-M.~Anton}
  \affiliation{The MacDiarmid Institute for Advanced Materials and Nanotechnology, School of Chemical and Physical Sciences, Victoria University of Wellington, PO Box 600, Wellington 6140, New Zealand}
\author{C.~Meyer}
  \affiliation{Institut N\'{e}el, Centre National de la Recherche Scientifique and Universit\'{e} Joseph Fourier, Bo\^{i}te Postale 166, F-38042 Grenoble Cedex, France}
\author{L.~Ranno}
  \affiliation{Institut N\'{e}el, Centre National de la Recherche Scientifique and Universit\'{e} Joseph Fourier, Bo\^{i}te Postale 166, F-38042 Grenoble Cedex, France}
\author{F.~Wilhelm}
  \affiliation{European Synchrotron Radiation Facility, Bo\^{i}te Postale 220, F-38043 Grenoble CEDEX, France}
\author{A.~Rogalev}
  \affiliation{European Synchrotron Radiation Facility, Bo\^{i}te Postale 220, F-38043 Grenoble CEDEX, France}

\begin{abstract}
Europium nitride is semiconducting and contains non-magnetic Eu$^{3+}$, but sub-stoichiometric EuN has Eu in a mix of 2+ and 3+ charge states. We show that at Eu$^{2+}$ concentrations near 15-20\% EuN is ferromagnetic with a Curie temperature as high as 120~K. The Eu$^{3+}$ polarization follows that of the Eu$^{2+}$, confirming that the ferromagnetism is intrinsic to the EuN which is thus a novel diluted magnetic semiconductor. Transport measurements shed light on the likely exchange mechanisms.
\end{abstract}

\date{\today}
\maketitle

Diluted magnetic semiconductors (DMSs), in which magnetic impurities are doped into a semiconducting host, offer important opportunities for use in spintronics technology as materials for spin injection or manipulation~\cite{Awschalom_Flatte,Zutic_DasSarma,Dietl_Ohno}. Understanding the exchange interactions in these systems is challenging, with a range of theoretical models proposed to describe the various systems~\cite{Coey_Venkatesan,Dietl_Ohno,Zunger_Raebiger,Coey_Paul}. The understanding is further complicated by the possible existence of magnetic impurity phases distinct from the semiconducting host, as these can often be of small enough dimensions to escape conventional detection methods~\cite{Ney,Shinde_Venkatesan,Chambers,Dietl_Wu}. Nevertheless, numerous examples of DMS systems have been reported, with ferromagnetic transition temperatures ranging from a few kelvin to far above room temperature~\cite{Sawicki_Bonanni,Dhar_Ploog,Cibert_Scalbert}. In the most well studied system, Mn-doped III-V semiconductors, the exchange mechanism is now reasonably well understood based on the modified Zener model of coupling mediated by  carriers~\cite{Dietl_Ohno}.

By contrast, it is relatively rare to find intrinsically ferromagnetic semiconductors, where the ordered magnetic moments are provided directly by the host cations~\cite{Nagaev,Mauger_Godart,Pappas_Fumagalli}. The most notable example is EuO~\cite{Mauger_Godart}, where the physics of the magnetic state in electron-doped samples remains controversial~\cite{Monteiro_Langridge,Liu_Tang}. The rare-earth nitride (REN) series, which are largely ionic with 3+ valence for the rare-earth and 3- for nitrogen, also contains such intrinsic ferromagnetic semiconductors, including GdN, DyN, and SmN~\cite{Leuenberger_Hessler,Granville_Trodahl,Preston_Lambrecht,Azeem_Kamba,Meyer_Trodahl,Larson_Schilfgaarde}. Europium nitride has also been demonstrated to be semiconducting~\cite{Richter_Lambrecht}, but EuN stands out amongst the RENs because the ground state of the Eu$^{3+}$ ion has configuration 4$f^6$ giving it a total angular momentum $J=0$, and thus it is non-magnetic~\cite{Johannes_Pickett}. Accordingly there are no ferromagnetic compounds based on Eu$^{3+}$. However, trivalent Eu does possess a non-zero spin angular momentum quantum number $S=3$ which has led to the suggestion that it might support ``hidden ferromagnetism''~\cite{Johannes_Pickett}. Furthermore, the first excited state $J=1$ lies close in energy to the ground state, so Eu$^{3+}$ has a relatively strong van Vleck susceptibility~\cite{VanVleck}. Thus, the magnetic properties of stoichiometric or doped EuN are of substantial interest.

We have previously demonstrated that epitaxial EuN films display a dominant paramagnetic signal that is at odds with that expected for a collection of Eu$^{3+}$ ions~\cite{Ruck_Meyer}. The origin of this signal was shown to be a concentration of a few percent of Eu$^{2+}$, most likely related to doping by nitrogen vacancies in the material~\cite{Richter_Lambrecht}. Furthermore, x-ray magnetic circular dichroism (XMCD) at the Eu L-edges showed that there is a partial polarization of the Eu$^{3+}$ that follows the Eu$^{2+}$ polarization~\cite{Ruck_Meyer}.

It is of fundamental interest to investigate the evolution of the magnetism in EuN as the quantity of Eu$^{2+}$ increases, thereby increasing the possibility of interaction between the localized magnetic moments. Here we present such a study based on EuN films with Eu$^{2+}$ concentrations as high as 15-20\%, and we show that such films are ferromagnetic at temperatures well above 100~K. XMCD results show that the Eu$^{2+}$ ions polarize the neighboring Eu$^{3+}$, showing that the ferromagnetism is not an artefact of an impurity phase and suggesting that  Eu$^{3+}$ plays a role in the exchange mechanism.

The 100-200~nm thick EuN films were grown onto substrates of either sapphire or GaN templates on sapphire by thermal evaporation of Eu in the presence of a flux of ionized nitrogen. In contrast to other rare-earth nitrides the use of an excited nitrogen source is essential for obtaining near-stoichiometric EuN films. The nitrogen partial pressure in the growth chamber was $3-5\times 10^{-4}$~mbar and the ions were accelerated through 125~V at a beam current of 0.37~mA. The films were grown at either room temperature or 680$^\circ$C, and were capped to prevent oxidation after growth using layers of either GaN or AlN. The films were characterized by reflection high-energy electron diffraction and x-ray diffraction (XRD), which showed that the 680$^\circ$C grown films are epitaxial with [111] orientation, while the room temperature grown films are polycrystalline with [111] texturing. There is no evidence in the XRD for any impurity phase, and all films show the expected lattice constant of 4.99~\AA~\cite{Richter_Lambrecht,Klemm_Winkelman,Brown_Clark}. As we will show below, the key difference between the 680$^\circ$C and the room temperature grown films is that the latter are more heavily doped and contain a substantially larger Eu$^{2+}$ concentration. We show representative data from films grown at the two temperatures, but the repeatability of the magnetization and transport results has been checked on additional samples grown under similar conditions.

The magnetization of the films was measured using a Quantum Design MPMS SQUID magnetometer. Further investigation of the magnetic state of the films was made by XMCD carried out at beam line ID12 of the European Synchrotron Radiation Facility in Grenoble, France. Measurements were made at grazing incidence in the total fluorescence yield detection mode, with the magnetic field applied in the film plane. Electrical transport measurements were conducted in a Quantum Design Physical Properties Measurement System using a four terminal geometry with contacts made using pressed indium.

\begin{figure}[h!]
  \centering
  \includegraphics[width=9cm]{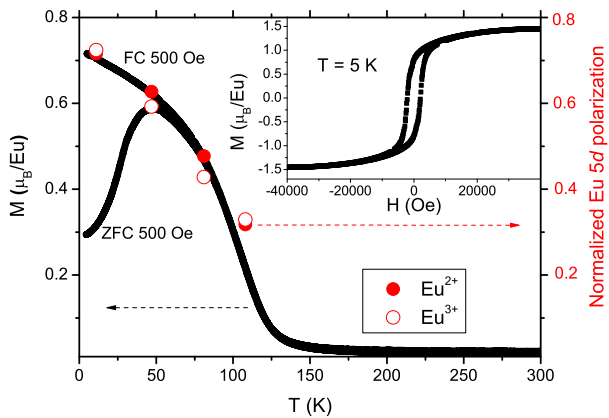}
  \caption{(color online) Field cooled (FC) and zero field cooled (ZFC) temperature dependent magnetization of a room temperature grown EuN film (solid lines) measured in a field of 500~Oe. Also shown are the Eu$^{2+}$ and Eu$^{3+}$ XMCD amplitudes (solid and open circles), which both follow the measured magnetisation. Inset: Hysteresis loop measured at 5~K.}
  \label{SQUID}
\end{figure}

In Figure~\ref{SQUID} we show our main result, namely that the room temperature grown films are ferromagnetic with a Curie temperature near 120~K as evidenced by the sharp rise in the temperature dependent magnetization. A clear hysteresis is observed at low temperatures (Fig.~\ref{SQUID} inset) along with saturation of the magnetization at around 1.4~$\mu_B$ per Eu ion. Assuming the magnetic response is associated with the Eu$^{2+}$ component of the film we estimate a rather large divalent fraction corresponding to about 20\% of the cations. A series of similar room temperature grown films all showed ferromagnetism, with Curie temperatures ranging from 100 to 120~K (estimated by extrapolating the steepest part of the magnetization curve back to zero). By contrast the 680$^\circ$C grown films display only a paramagnetic response whose magnitude is consistent with Eu$^{2+}$ concentrations of around 2-5\%.

\begin{figure}[h!]
  \centering
  \includegraphics[width=9cm]{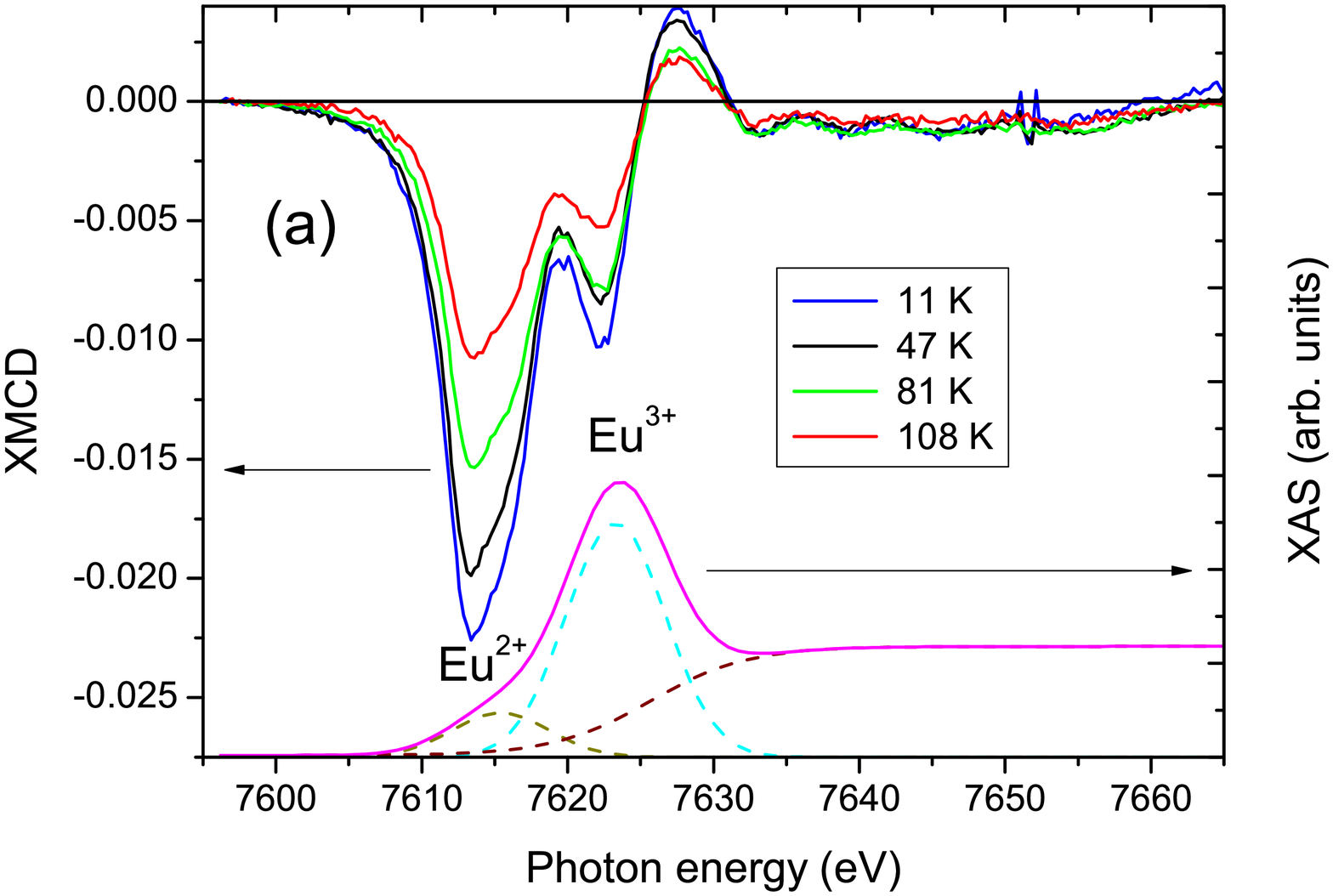}
  \includegraphics[width=9cm]{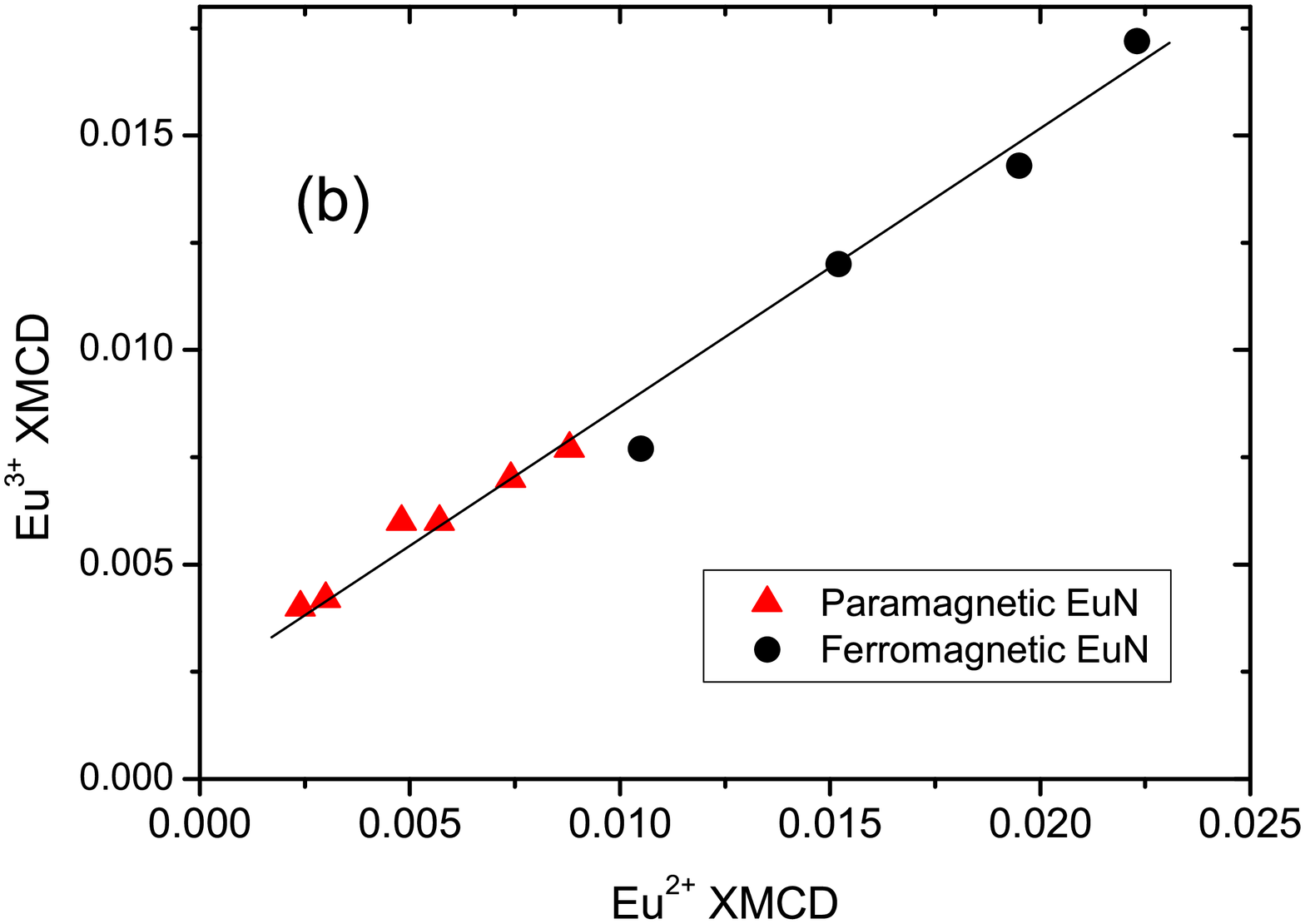}
  \caption{(color online) (a)~Eu L$_2$-edge x-ray absorption and XMCD at various temperatures from ferromagnetic EuN. The fit to the absorption spectrum (dashed colored lines) implies that about 15\% of the Eu ions are in the 2+ charge state. Strong XMCD with similar temperature dependence is observed for both the Eu$^{2+}$ and Eu$^{3+}$ features. (b)~Eu$^{3+}$ versus Eu$^{2+}$ polarization of the $5d$ electrons extracted from the XMCD spectra. The black symbols are from the ferromagnetic film in (a), the red symbols from the paramagnetic film reported in Ref.~[\onlinecite{Ruck_Meyer}]. The solid line is a guide to the eye.}
  \label{XMCD}
\end{figure}

To understand the origin of the ferromagnetism we have carried out XMCD on a ferromagnetic EuN film grown at room temperature. The L$_{2,3}$-edge XMCD involves the transition $2p^6 5d^0 \rightarrow 2p^5 5d^1$ so it interrogates the polarization of the $5d$ empty-state conduction band orbitals. The x-ray absorption spectrum at the L$_2$-edge shown in Figure~\ref{XMCD}(a) shows a clear shoulder at 7615~eV superimposed on the usual Eu$^{3+}$ white line absorption centered at 7624~eV. Atomic multiplet calculations clearly identify the shoulder as originating from absorption by Eu$^{2+}$ ions~\cite{Richter_Lambrecht,Thole_Esteva}. This feature is substantially stronger than the corresponding shoulder seen in a paramagnetic epitaxial EuN film~\cite{Ruck_Meyer}, confirming the much larger Eu$^{2+}$ concentration in the room temperature grown films. The curve fitting of the 2+ and 3+ peaks shown in the figure implies a Eu$^{2+}$ concentration of around 15\%, consistent within uncertainty with the value extracted above from the saturation magnetization.

The corresponding XMCD spectra taken at various temperatures in a field of 3~T are also plotted in Fig.~\ref{XMCD}(a). The strongest feature near 7615~eV is clearly associated with Eu$^{2+}$. The strength of this XMCD feature at the lowest temperature is roughly three times stronger than in the paramagnetic epitaxial films~\cite{Ruck_Meyer}, and its temperature dependence follows closely the measured magnetization as shown by the solid symbols in Fig.~\ref{SQUID} (the disagreement between the SQUID and XMCD amplitudes at 105~K is a result of the much higher measurement field used for XMCD). These observations confirm the origin of the ferromagnetism to be the Eu in the film.

The remainder of the XMCD features are associated mostly with Eu$^{3+}$, and interestingly these also show a strong signature of the ferromagnetism. Similar to the Eu$^{2+}$ XMCD the amplitude of the 3+ signal is much larger than that seen in paramagnetic films at similar applied fields. Furthermore, rather than following a van Vleck temperature dependence, the Eu$^{3+}$ signal closely follows the Eu$^{2+}$ signal (open symbols in Fig.~\ref{SQUID}), implying that there is a strong exchange coupling between the Eu$^{2+}$ and Eu$^{3+}$ ions~\cite{Matsuda_Wada}. This is further demonstrated in Figure~\ref{XMCD}(b) which shows the Eu$^{3+}$ polarization plotted against the Eu$^{2+}$ polarization determined from the XMCD using the method described in Ref.~[\onlinecite{Ruck_Meyer}]. The red triangles represent data from the paramagnetic sample of Ref.~[\onlinecite{Ruck_Meyer}]. The black circles, representing the ferromagnetic sample, show much larger polarization for both species but they follow the same trend as the paramagnetic sample implying that the coupling between the Eu$^{2+}$ and Eu$^{3+}$ ions is of the same nature in each case, with the key difference simply being the concentration. We stress that the strong coupling between the Eu$^{2+}$ and Eu$^{3+}$ is compelling evidence that the ferromagnetic phase is not simply an impurity, such as electron-doped EuO~\cite{Mauger_Godart,Liu_Tang}, but rather represents the response of the EuN matrix containing a large concentration of Eu$^{2+}$ ions.

\begin{figure}[h!]
  \includegraphics[width=8cm]{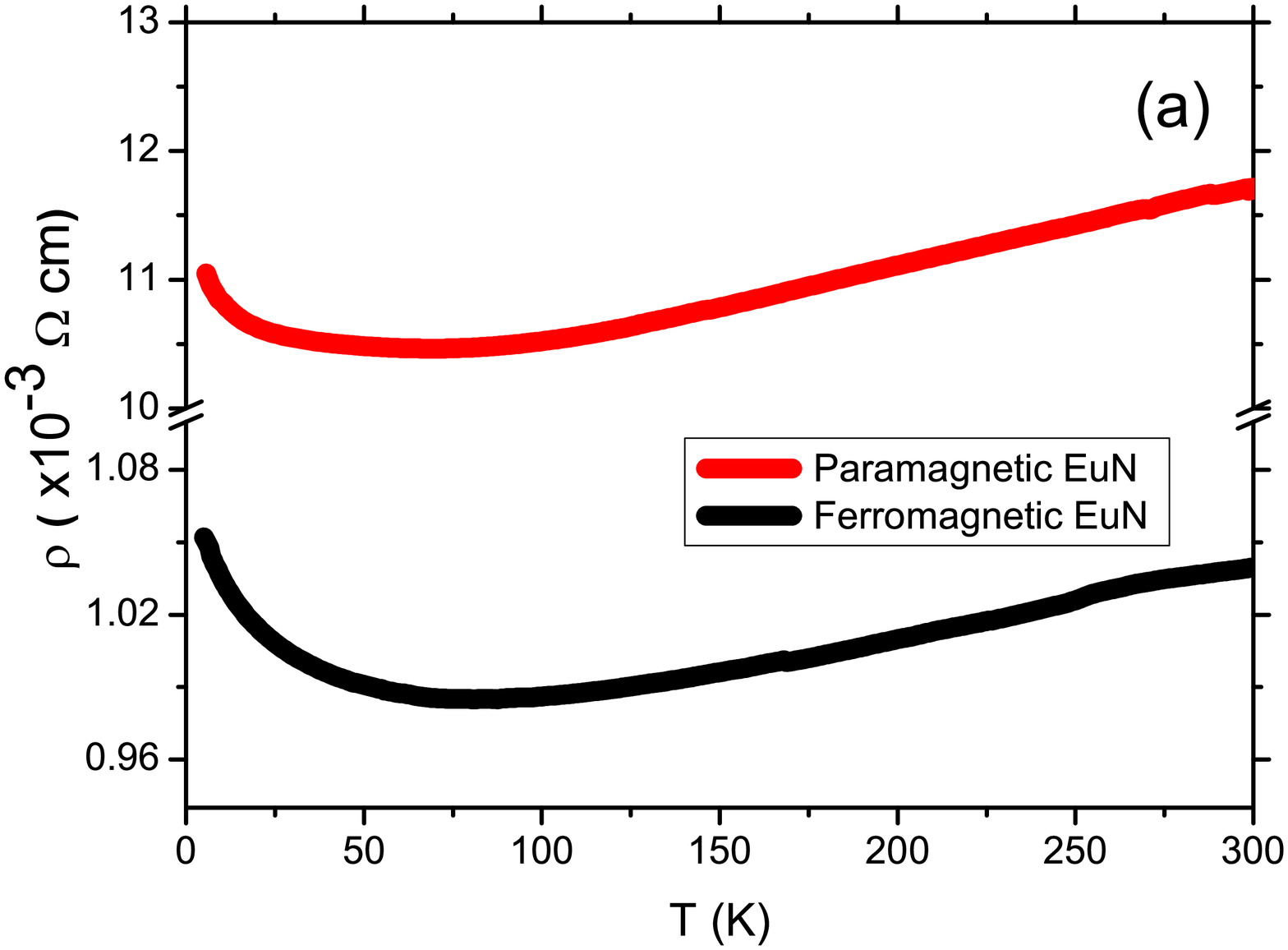}
  \includegraphics[width=8cm]{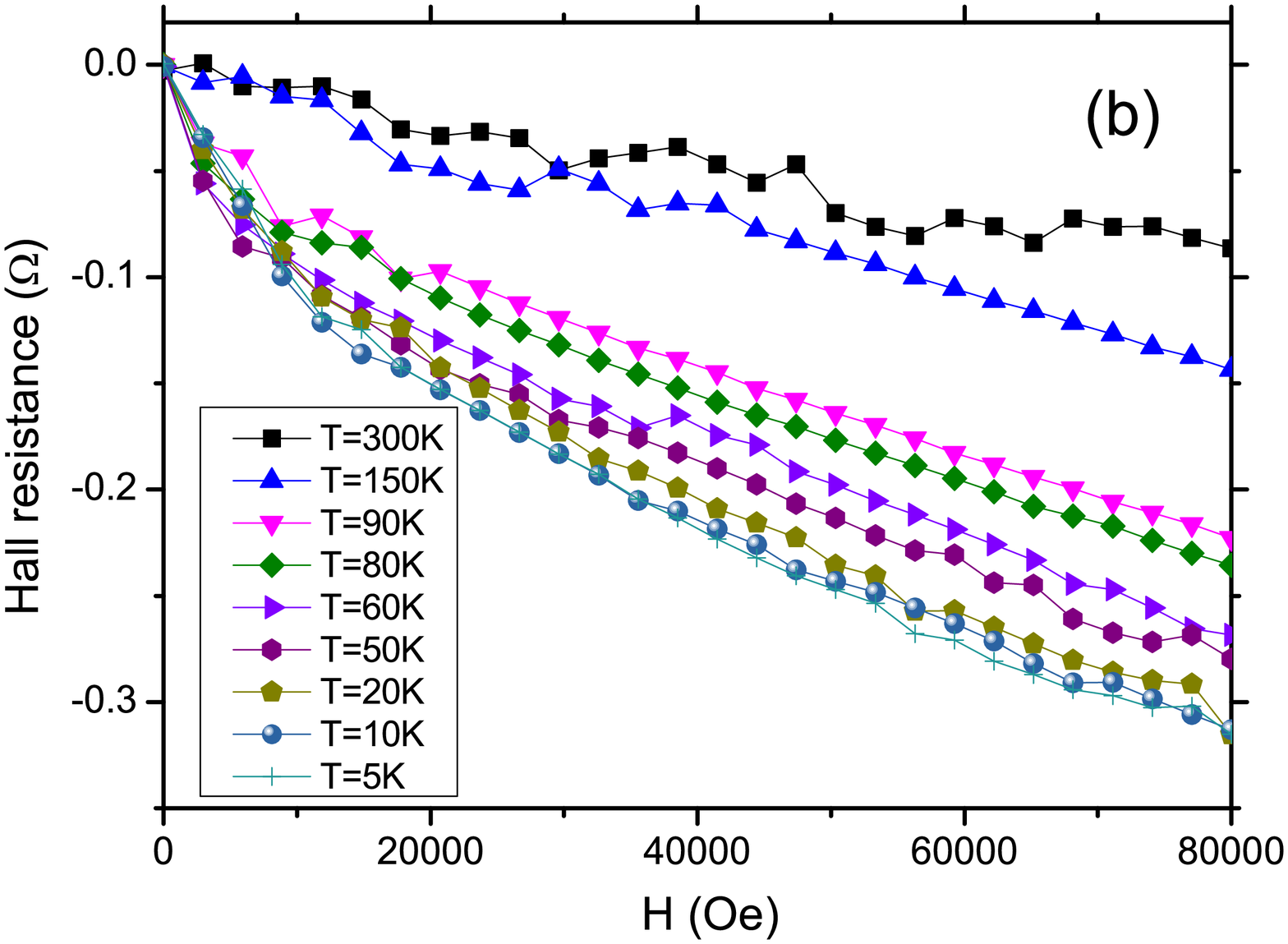} 
  \includegraphics[width=8cm]{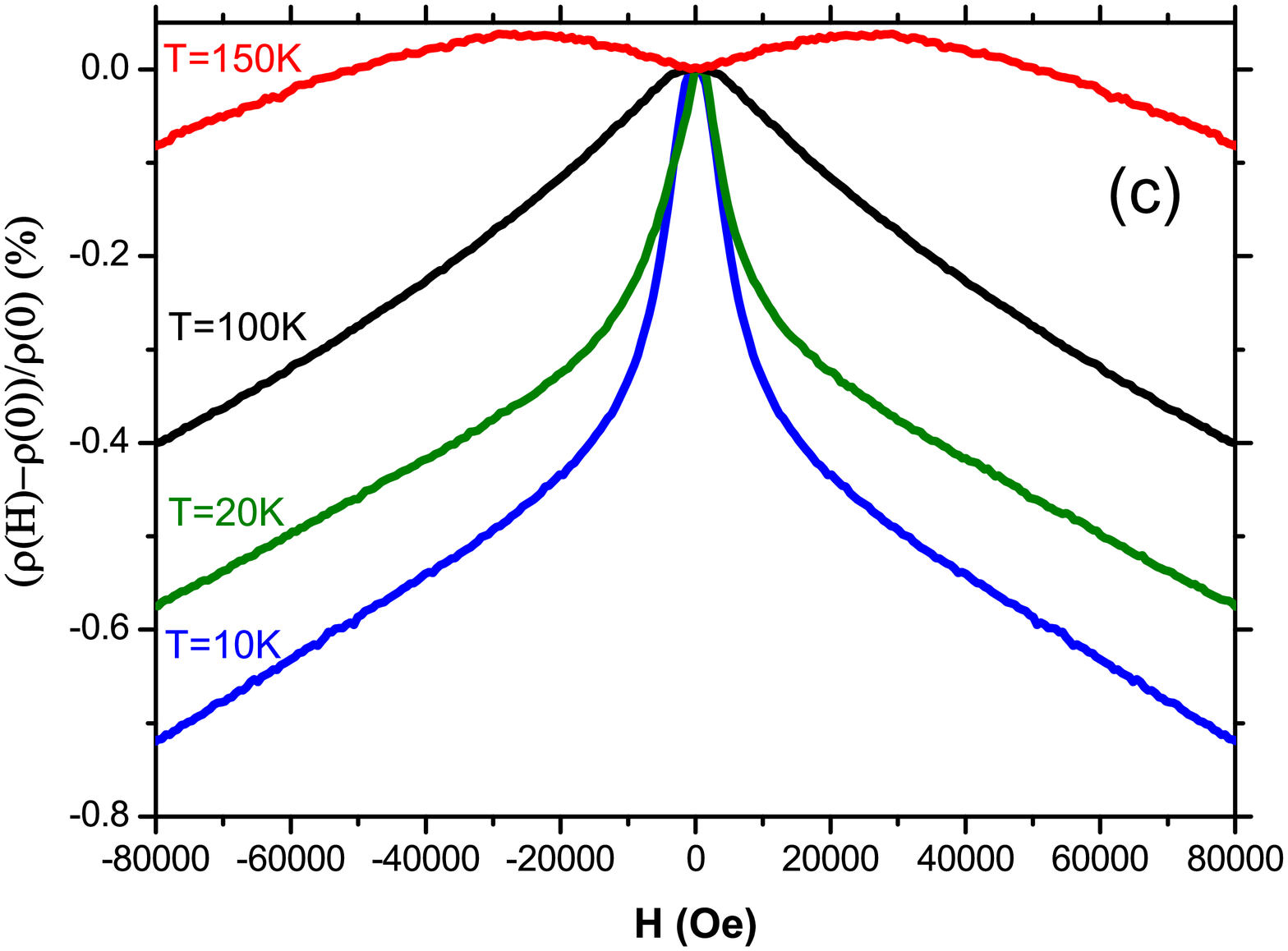} 

  \caption{(color online) (a)~Temperature dependent resistivity of ferromagnetic (black) and paramagnetic (red) EuN. (b)~Hall resistance of ferromagnetic EuN, showing an anomalous Hall effect below the T$_C$ of 120~K. (c)~Magnetoresistance of ferromagnetic EuN showing cusp-like behavior at low fields for temperatures below T$_C$.}
  \label{Transport}
\end{figure}

We have further investigated the source of the doping and the nature of the exchange mechanism that couples the Eu$^{2+}$ by measuring the transport properties of the films. Figure~\ref{Transport}(a) shows the temperature dependent resistivity of a room temperature grown ferromagnetic film and a paramagnetic film grown at 680$^\circ$C. The paramagnetic film shows a metallic temperature dependence at high temperature, developing a negative temperature coefficient of resistance below about 60~K. The magnitude of the resistivity is rather high ($\approx11~\mathrm{m}\Omega$cm), consistent with the conclusion that these EuN films are semiconductors doped to degeneracy by a high concentration of nitrogen vacancies. This is further supported by Hall effect measurements that give a carrier concentration at room temperature of $-8\times10^{20}$~cm$^{-3}$ (i.e., the carriers are electrons). The origin of the upturn in the resistivity below 60~K is uncertain. Magnetic scattering from the Eu$^{2+}$ in the film could lead to a Kondo effect~\cite{Kondo,He_Wang}, and indeed the resistivity does follow the expected logarithmic temperature dependence.

The room temperature grown ferromagnetic film shows a qualitatively similar temperature dependent resistivity, although the magnitude is substantially smaller and the carrier concentration is larger ($4\times10^{21}$~cm$^{-3}$). The ratio of carrier concentration between the two films is similar to the ratio of Eu$^{2+}$ content, indicating a link between the two quantities. The low temperature resistivity upturn occurs at higher temperature in the ferromagnetic sample than the paramagnetic sample, and it constitutes a larger fractional change in resistivity in the film with more Eu$^{2+}$. However, the ferromagnetic film has a larger mobility (1.5~cm$^2$V$^{-1}$s$^{-1}$) than the paramagnetic film (0.7~cm$^2$V$^{-1}$s$^{-1}$), supporting the conclusion that the upturn is related to magnetic scattering rather than weak localization~\cite{Altshuler_Larkin}.

There is no sharp anomaly at the Curie temperature in the ferromagnetic sample, as is often observed in ferromagnets where it can be caused either by scattering from magnetic fluctuations~\cite{deGennes_Friedel,Fisher_Langer,He_Wang} or by a change in carrier concentration as the sample enters the ferromagnetic state~\cite{Mauger_Godart,Leuenberger_Hessler,Granville_Trodahl}. On the other hand evidence for the magnetic ordering is clearly seen in the form of an anomalous Hall effect that sets in below $T_C$ [Fig.~\ref{Transport}(b)], the strength of which is enhanced by the relatively large resistivity in these films. Evidence for coupling between the magnetic order and the electrical transport is also present in the magnetoresistance presented in Fig.~\ref{Transport}(c). It shows a negative parabolic behavior above $T_C$, characteristic of scattering from uncorrelated magnetic impurities~\cite{BealMonod_Weiner}, with an additional positive contribution evident at low field~\cite{Dietl_Wu}. These features disappear below $T_C$ to be replaced by a sharp negative cusp at low fields followed by a near-linear high-field behavior, similar to the behavior observed in other ferromagnets~\cite{Csontos_Mihaly}. By contrast, the magnetoresistance of the paramagnetic sample is parabolic down to low temperature with a small cusp observable only below 10~K. Similarly, the paramagnetic films show evidence for an anomalous Hall effect only below 10~K where the Eu$^{2+}$ becomes strongly polarized in the large measurement fields.

Based on the evidence presented above and previous calculations and measurements of the electronic structure of stoichiometric EuN we propose a simple model for the formation of Eu$^{2+}$ in EuN. The underlying band structure is semiconducting, but with the Eu$^{2+}$ $4f^7$ ($^8S$) level lying very close to the bottom of the conduction band~\cite{Richter_Lambrecht}. The presence of large quantities of nitrogen vacancies shifts the Fermi level into the conduction band, and at carrier concentrations above $\sim10^{20}$~cm$^{-3}$ it approaches the $^8S$ level that thus becomes populated. This is similar to a proposed model of the electronic structure of sub-stoichiometric YbN~\cite{Degiorgi_Wachter}, although there the Yb$^{2+}$ is nonmagnetic and there is no magnetic ordering.

Given the above model it is interesting to seek evidence for an enhancement of the effective mass in the heavily doped samples where the Fermi level approaches the Eu $^8S$ level. To investigate this possibility we write the resistivity as the sum of a phonon contribution ($\rho_{\mathrm{ph}}$) and a contribution from disorder scattering involving both lattice defects and magnetic inhomogeneity ($\rho_{\mathrm{dis}}$):

          \begin{equation}\label{resmod}
          \rho(T)=\rho_{\mathrm{dis}}+\rho_{\mathrm{ph}}=\frac{m^*}{ne^2\tau_{\mathrm{dis}}}+\frac{m^*}{ne^2\tau_{\mathrm{ph}}},
          \end{equation}

\noindent where $m^*$ is the carrier effective mass, $n$ is the carrier concentration, $e$ is the electron's charge, $\tau_{\mathrm{ph}}$ is the phonon scattering time, and $\tau_{\mathrm{dis}}$ is the combined magnetic and quenched disorder scattering time. At high temperature $\tau_{\mathrm{dis}}$ is temperature independent and the phonon scattering rate $\tau_{\mathrm{ph}}^{-1}=cT$ with $c$ a constant, so we can express the effective mass as

\begin{equation}
m^* = \frac{1}{ne^2c}\frac{d\rho}{dT}.
\end{equation}

\noindent Assuming the phonon scattering rate, and hence the constant $c$, is the same in all samples, and using the measured resistivity slopes and carrier concentrations, we can seek variations in $m^*$ between samples. Doing so we find that the paramagnetic film in Fig.~\ref{Transport}(a) has a larger $m^*$ than the ferromagnetic film by a factor of nearly three. This is the same within the uncertainty among all of the films, and we see no evidence for a systematic variation in $m^*$ with Eu$^{2+}$ concentration. Once again this is consistent with conclusions obtained from YbN~\cite{Degiorgi_Wachter}.

Next, we consider the possible exchange interactions present in the films. The carrier concentration in the ferromagnetic samples is larger than in the paramagnetic films, suggesting that carrier mediated mechanisms may play an important role. Indeed, the conduction band states are formed primarily from Eu 5$d$ orbitals, and the XMCD results show a very clear polarization of these states. This is similar to the polarization of Eu$^{3+}$ seen in the mixed valence compounds EuNi$_2$P$_2$ and EuNi$_2$(Si$_{0.18}$Ge0.82)$_2$, although there the polarization was induced by a very large applied field~\cite{Matsuda_Wada}.  At the large Eu$^{2+}$ concentrations where ferromagnetism occurs there will be many nearest-neighbor Eu$^{2+}$ ions on the cation sublattice, allowing for short-ranged exchange interactions. This will naturally lead to a percolating type of magnetic ordering nucleating at regions of high Eu$^{2+}$ density, and this percolating nature might explain the lack of a cusp at T$_C$ in the temperature dependent resistivity. Finally, we note that the underlying matrix of Eu$^{3+}$ ions is also polarizable due to the small energy gap to the $J=1$ excited state, which could lead to a Van Vleck type contribution to the exchange interaction as has been reported for Cr-doped Bi$_2$Sb$_3$~\cite{Yu_Fang}. XMCD measurements at the Eu M-edge would be of interest to probe the 4$f$ levels directly.

In summary, we have shown that EuN with a large fraction of the Eu ions in the 2+ charge state is ferromagnetic at temperatures as high as 120~K. It thus represents a novel dilute magnetic semiconducting system, with the magnetism contributed largely by the Eu$^{2+}$, but where the host lattice based on Eu$^{3+}$ is also polarizable. The concentration of Eu$^{2+}$ ions is correlated with the charge carrier concentration, allowing us to propose a simple model for the formation of Eu$^{2+}$. The large concentration of Eu$^{2+}$ in the ferromagnetic samples requires that many are nearest neighbors on the cation lattice allowing for short-ranged exchange interactions which may be supported by interactions involving the charge carriers and also even the polarizable Eu$^{3+}$ background. The relatively simple physical structure of this system may make it an attractive testing ground for theories of exchange interactions in diluted magnetic systems.

\begin{acknowledgments}
We acknowledge funding from the NZ Foundation for Research, Science, and Technology (VICX0808) and the Marsden Fund (08-VUW-030). The MacDiarmid Institute is supported by the New Zealand
Centres of Research Excellence Fund. We are grateful to Walter Lambrecht for helpful discussions.
\end{acknowledgments}


\end{document}